\documentclass[twocolumn,groupedaddress,superscriptaddress,nofootinbib,amsmath,amssymb]{revtex4}
\usepackage{amsbsy,subfigure,hyperref,bbm,times}
\usepackage[T1]{fontenc}
\usepackage{graphicx}
\usepackage{dcolumn}
\usepackage{bm}

\pdfoutput=1

\newcommand{\ket}[1]{|#1\rangle}

\providecommand{\openone}{\leavevmode\hbox{\small1\kern-3.8pt\normalsize1}}

\begin{document}

\title{Entanglement of photons in their dual wave-particle nature}

\author{Adil S. Rab}
\affiliation{Dipartimento di Fisica, Sapienza Universit\`{a} di Roma, Piazzale Aldo Moro, 5, I-00185 Roma, Italy}

\author{Emanuele Polino}
\affiliation{Dipartimento di Fisica, Sapienza Universit\`{a} di Roma, Piazzale Aldo Moro, 5, I-00185 Roma, Italy}

\author{Zhong-Xiao Man} 
\email{zxman@mail.qfnu.edu.cn}
\affiliation{Shandong Provincial Key Laboratory of Laser Polarization and Information Technology, Department of Physics, Qufu Normal University, Qufu 273165, China}

\author{Nguyen Ba An}
\affiliation{Center for Theoretical Physics, Institute of Physics, Vietnam Academy of Science and Technology (VAST), 18
Hoang Quoc Viet, Cau Giay, Hanoi, Vietnam}

\author{Yun-Jie Xia}
\affiliation{Shandong Provincial Key Laboratory of Laser Polarization and Information Technology, Department of Physics, Qufu Normal University, Qufu 273165, China}

\author{Nicol\`{o} Spagnolo}
\affiliation{Dipartimento di Fisica, Sapienza Universit\`{a} di Roma, Piazzale Aldo Moro, 5, I-00185 Roma, Italy}

\author{Rosario Lo Franco} 
\email{rosario.lofranco@unipa.it}
\affiliation{Dipartimento di Energia, Ingegneria dell'Informazione e Modelli Matematici, Universit\`{a} di Palermo, Viale delle Scienze, Edificio 9, 90128 Palermo, Italy}
\affiliation{Dipartimento di Fisica e Chimica, Universit\`a di Palermo, via Archirafi 36, 90123 Palermo, Italy}

\author{Fabio Sciarrino}
\email{fabio.sciarrino@uniroma1.it}
\affiliation{Dipartimento di Fisica, Sapienza Universit\`{a} di Roma, Piazzale Aldo Moro, 5, I-00185 Roma, Italy}

\maketitle

\textbf{Wave-particle duality is the most fundamental description of the nature of a quantum object which behaves like a classical particle or wave depending on the measurement apparatus \cite{Wheeler1,MaRMP,Wheeler2,manning}. 
On the other hand, entanglement represents nonclassical correlations of composite quantum systems \cite{horodecki2009quantum,lofranco2015quantum,BrunnerReview}, being also a key resource in quantum information \cite{vedralReview,laddReview,Wang16}. 
Despite the very recent observations of wave-particle superposition~\cite{quan-delay-exp,
delay-exp1,delay-exp2,delay-exp3,delay-exp4,delay-exp5,pho-test,Ma22012013} and entanglement \cite{BrunnerReview,hensenBell,PhysRevLett.115.250401,PhysRevLett.115.250402,Hand17}, whether these two fundamental traits of quantum mechanics can emerge simultaneously remains an open issue. 
Here we introduce and experimentally realize a scheme that deterministically generates wave-particle entanglement of two photons. The elementary tool allowing this achievement is a scalable single-photon setup which can be in principle extended to generate multiphoton wave-particle entanglement.
Our study reveals that photons can be entangled in their dual wave-particle nature and opens the way to potential
applications in quantum information protocols exploiting the wave-particle degrees of freedom to encode qubits.}

\begin{figure}[b!]
	\centering
	\includegraphics[width= 0.5\textwidth]{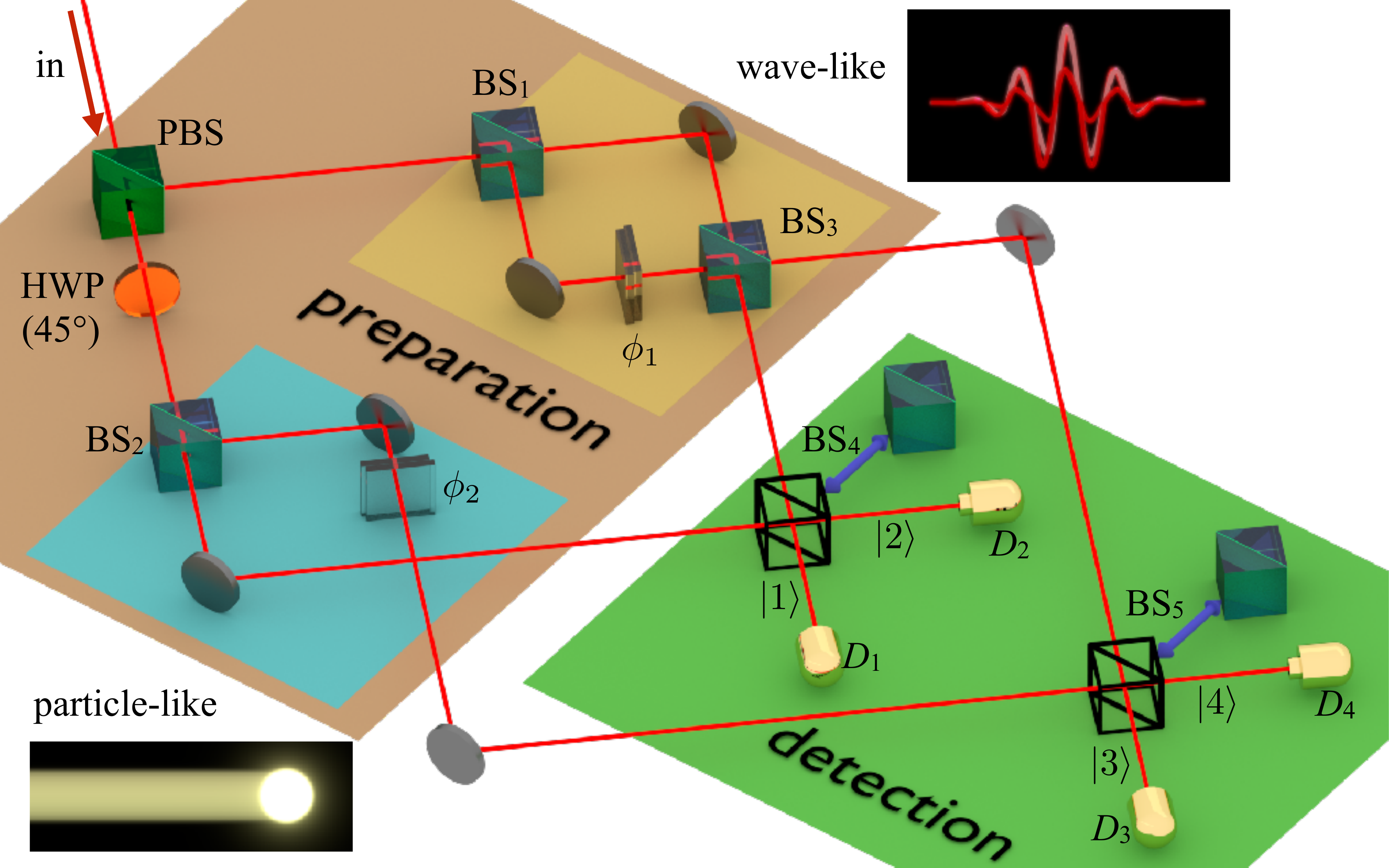} 
	\caption{{\bf Conceptual figure of the wave-particle toolbox.} A single photon is coherently separated in two spatial modes by means of a polarizing beam-splitter (PBS) according to its initial polarization state (in). A half-wave plate (HWP) is placed after the PBS to obtain equal polarizations between the two modes. One mode is injected in a complete Mach-Zehnder interferometer (MZI) with phase $\phi_{1}$, thus exhibiting wave-like behavior. The second mode is injected in a Mach-Zehnder interferometer lacking the second beam-splitter, thus exhibiting particle-like behavior (no dependence on $\phi_{2}$). The output modes are recombined on two symmetric beam-splitters (BS$_{4}$, BS$_{5}$), which can be removed to change the measurement basis. Detectors ($D_{1}$, $D_{2}$, $D_{3}$, $D_{4}$) are placed on each final path ($\ket{1}$, $\ket{2}$, $\ket{3}$, $\ket{4}$).}
		\label{fig:figure1}
\end{figure}

Quantum mechanics is one of the most successful theories in describing atomic-scale systems albeit its properties remain bizarre and counterintuitive from a classical perspective. A paradigmatic example is the wave-particle duality of a single quantum system, which can behave like both particle and wave to fit the demands of the experiment's configuration \cite{Wheeler1}. This double nature is well reflected by the superposition principle and evidenced for light by Young-type double-slit experiments \cite{MaRMP,walbornDoubleSlit}, where single photons from a given slit can be detected (particle-like behavior) and interference fringes observed (wave-like behavior) on a screen behind the slits. A double-slit experiment can be simulated by sending photons into a Mach-Zehnder interferometer (MZI) via a semitransparent mirror (beam splitter) \cite{MaRMP,walbornDoubleSlit}. A representative experiment with MZI, also performed with a single atom \cite{manning}, is the Wheeler's delayed-choice (WDC) experiment \cite{Wheeler1,Wheeler2}, where one can choose to observe the particle or wave character of the quantum object after it has entered the interferometer. These experiments rule out the existence of some extra information hidden in the initial state telling the quantum object which character to exhibit before reaching the measurement apparatus. Very recent quantum WDC experiments, using quantum detecting devices and requiring ancilla photons or post-selection, have then shown that wave and particle behaviors of a single photon can coexist simultaneously, with a continuous morphing between them \cite{quan-delay-exp,delay-exp1,delay-exp2,delay-exp3,delay-exp4,delay-exp5,pho-test}.

\begin{figure*}[tbp]
\centering
\includegraphics[width=\textwidth]{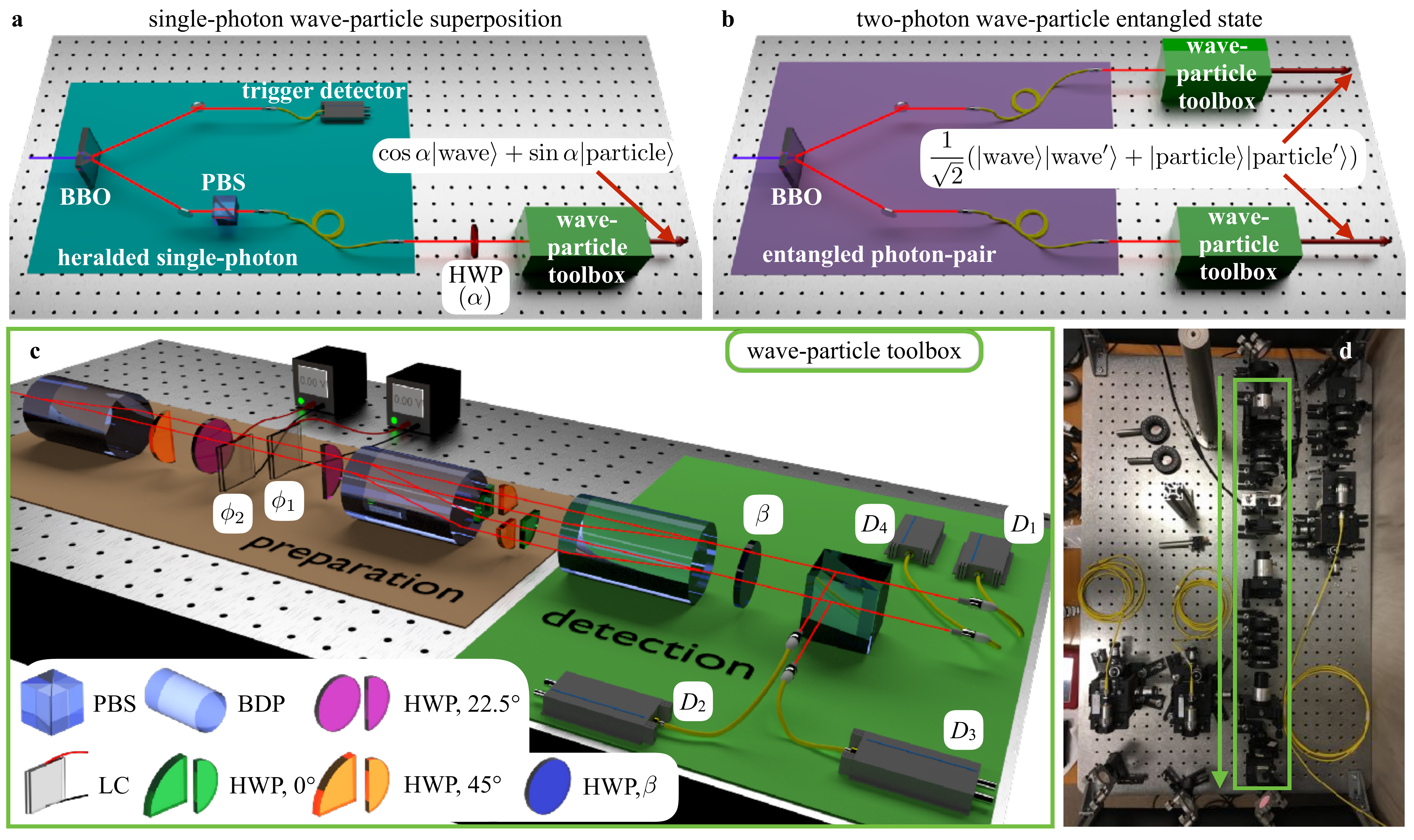}
\caption{{\bf Experimental setup for wave-particle states.} {\bf a}, Overview of the apparatus for the generation of single-photon wave-particle superposition. An heralded single-photon is prepared in an arbitrary linear polarization state through an half-wave plate rotated at an angle $\alpha/2$ and injected into the wave-particle toolbox. {\bf b}, Overview of the apparatus for the generation of a two-photon wave-particle entangled state. Each photon of a polarization entangled state is injected into an independent wave-particle toolbox to prepare the output state. {\bf c}, Actual implemented wave-particle toolbox, reproducing the action of the scheme shown in Fig. \ref{fig:figure1}. The interferometer is composed of beam-displacing prisms (BDP), half-wave plates (HWP), and liquid crystal devices (LC), the latters changing the phases $\phi_{1}$ and $\phi_{2}$. The output modes are finally separated by means of a polarizing beam-splitter (PBS). The scheme corresponds to the presence of BS$_4$ and BS$_5$ in Fig. \ref{fig:figure1} for $\beta=22.5^{\circ}$, while setting $\beta = 0$ equals to the absence of BS$_4$ and BS$_5$. {\bf d}, Picture of the experimental apparatus. The green frame highlights the wave-particle toolbox.}
\label{fig:figure2}
\end{figure*}

When applying the superposition principle to composite systems, another peculiar quantum feature arises, namely the entanglement among degrees of freedom of the constituent particles (e.g., spins, energies, spatial modes, polarizations) \cite{horodecki2009quantum,lofranco2015quantum}. Entanglement gathers fundamental quantum correlations among particle properties which are at the core of nonlocality \cite{BrunnerReview,hensenBell,PhysRevLett.115.250401,PhysRevLett.115.250402,Hand17} and exploited as essential ingredient for developing quantum technologies \cite{vedralReview,laddReview,Wang16}. Superposition principle and entanglement have been amply debated within classical-quantum border, particularly whether macroscopically distinguishable states (i.e., distinct quasiclassical wave packets) of a quantum system could be prepared in superposition states \cite{harocheNobelLect}. While superpositions of coherent states of a single quantum system (also known as ``cat states'' from the well-known Schr\"odinger's epitome) have been observed for optical or microwave fields starting from two decades ago \cite{harocheNobelLect,brunePRL,Ourjoumtsev,Deleglise,Vlastakis607}, the creation of entangled coherent states of two separated subsystems has remained a demanding challenge, settled only very recently by using superconducting microwave cavities and Josephson junction-based artificial atoms \cite{Wang1087}. An analogous situation exists in the context of wave-particle duality where, albeit wave-particle superpositions of a photon have been reported \cite{quan-delay-exp,delay-exp1,delay-exp2,delay-exp3,delay-exp4,delay-exp5,pho-test}, entangled states of photons correlated in their wave-particle degrees of freedom are still unknown. 

In this work we experimentally demonstrate that wave-particle entanglement of two photons is achievable deterministically. We reach this goal by introducing and doubling a scalable all-optical scheme which is capable to generate, in an unconditional manner, controllable single-photon wave-particle superposition states. Parallel use of this basic toolbox then allows the creation of multiphoton wave-particle entangled states. 

\begin{figure*}[tbp]
\centering
\includegraphics[width=\textwidth]{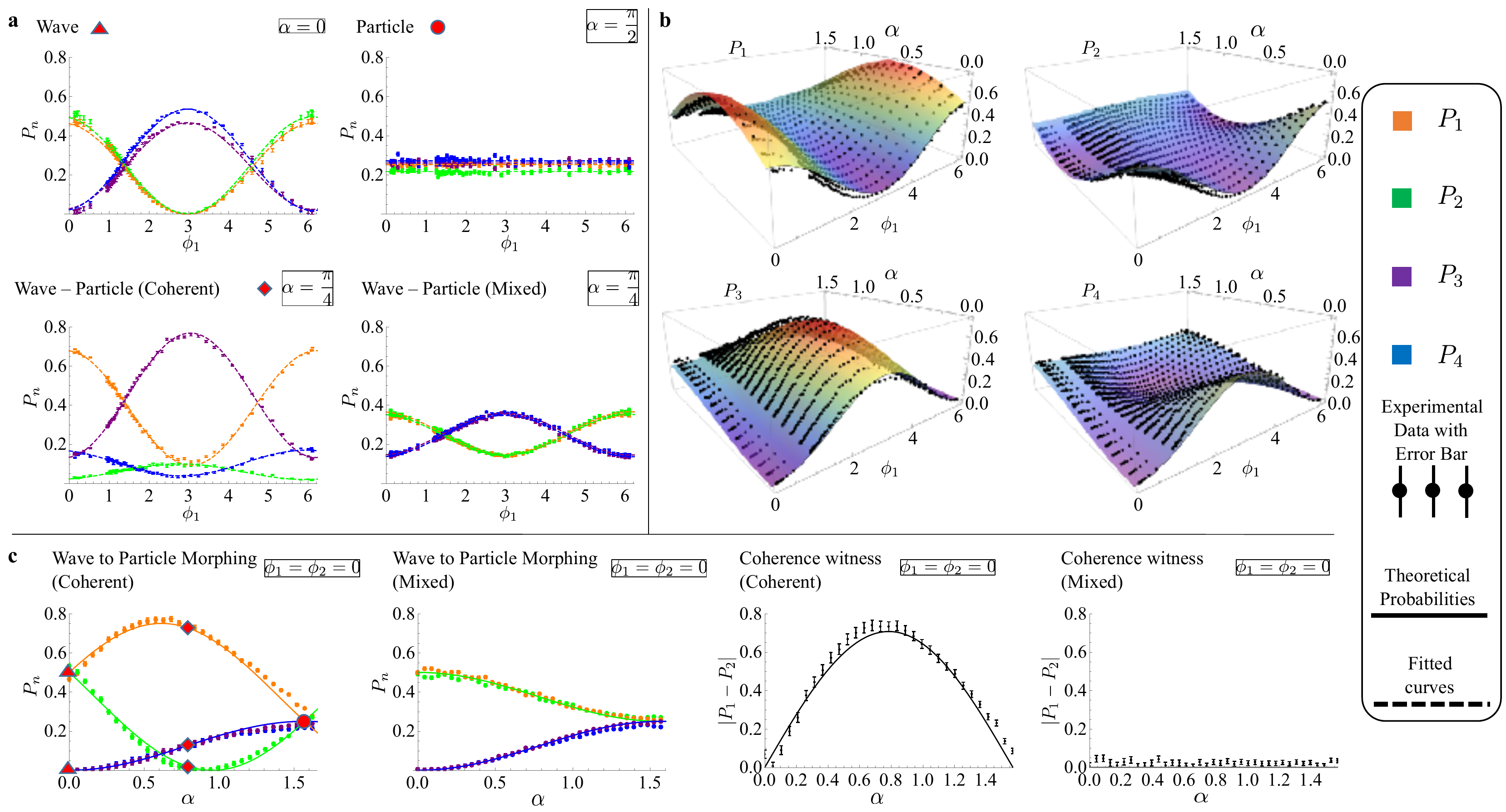}
\caption{{\bf Generation of wave-particle superposition with a single-photon state.} {\bf a}, Measurements of the output probabilities $P_{n}$ as a function of the phase $\phi_{1}$, for different values of $\alpha$. In clock-wise order: wave behavior ($\alpha=0$), particle behavior ($\alpha=\pi/2$), incoherent mixture of wave and particle behaviors ($\alpha=\pi/4$), and coherent wave-particle superposition ($\alpha=\pi/4$).  {\bf b}, 3d plots output probabilities $P_{n}$ as a function of the phase $\phi_{1}$ and of the angle $\alpha$.  {\bf c}, Evidence of the generation of wave-particle superpositions. From left to right: probabilities $P_{n}$ as a function of $\alpha$ in the coherent case and for an incoherent mixture, witness $\mathcal{W}_C = \vert P_{1} - P_{2} \vert$ in the coherent case and for an incoherent mixture (the latter showing no dependence on $\alpha$). Points: experimental data. Solid curves and surfaces: theoretical expectations. Dashed curves: best-fit of the experimental data. Error bars are due to the Poissonian statistics of single photon counting.}
\label{fig:figure3}
\end{figure*}

\textbf{Single-photon toolbox.} The theoretical sketch of the wave-particle scheme for the single photon is displayed in Fig.~\ref{fig:figure1}. A photon is initially prepared in a polarization state $\left| \psi _{0}\right\rangle =\cos \alpha \left| V\right\rangle +\sin \alpha \left| H\right\rangle$, where $\left| V\right\rangle$ and $\left| H\right\rangle$ are the
vertical and horizontal polarization states and $\alpha$ is adjustable by a preparation half-wave plate (not shown in the figure). After crossing the setup with beam-splitter BS$_4$ and BS$_5$ inserted (see Supplementary Information for details), the photon state is
\begin{equation} \label{wp-state}
\left| \psi _{f}\right\rangle =\cos \alpha \left| \mathrm{wave}\right\rangle
+\sin \alpha \left| \mathrm{particle}\right\rangle, 
\end{equation}
where the states
\begin{eqnarray}\label{wave-particle}
&\left| \mathrm{wave}\right\rangle =\frac{e^{i\frac{\phi _{1}}{2}}}{\sqrt{2}}\left[\cos\frac{\phi _{1}}{2}(\left| 1\right\rangle +\left| 2\right\rangle)
-i\sin\frac{\phi _{1}}{2}(\left| 3\right\rangle +\left| 4\right\rangle)\right], & \nonumber \\
&\left| \mathrm{particle}\right\rangle =\frac{1}{2}(\left| 1\right\rangle
-\left| 2\right\rangle +e^{i\phi _{2}}\left| 3\right\rangle -e^{i\phi
_{2}}\left| 4\right\rangle ),& 
\end{eqnarray}
operationally represent the capacity ($\ket{\mathrm{wave}}$) and incapacity ($\ket{\mathrm{particle}}$) of the photon to produce interference \cite{quan-delay-exp,delay-exp5}. In fact, for the $\left| \mathrm{wave}\right\rangle$
state the probability of detecting the photon in the path $\ket{n}$ ($n = 1,2,3,4$) depends on the phase $\phi _{1}$: the photon must have traveled along both paths simultaneously (see upper MZI in Fig.~\ref{fig:figure1}), revealing its wave nature.
Instead, for the $\left| \mathrm{particle}\right\rangle $ state the
probability that the photon is detected in a certain path is always $1/4$, 
regardless of phase $\phi _{2}$: thus, the photon must have crossed only one of the two paths (see lower MZI of Fig.~\ref{fig:figure1}), showing its particle nature. Notice that the scheme is designed in such a way that $\ket{V}$ ($\ket{H}$) leads to the $\ket{\mathrm{wave}}$ ($\ket{\mathrm{particle}}$) state.

From equation~(\ref{wp-state}), the probability $P_{n}$ of detecting the
photon along path $\ket{n}$ is expected to depend on all the involved parameters, $P_{n}=P_{n}(\alpha ,\phi_{1},\phi _{2})$. These probabilities are 
\begin{equation}\label{Pn}
P_{1} = \mathcal{P}_c + \mathcal{I}_c,\
P_{2} =\mathcal{P}_c - \mathcal{I}_c, \
P_{3} =\mathcal{P}_s + \mathcal{I}_s, \ 
P_{4} =\mathcal{P}_s - \mathcal{I}_s, 
\end{equation}
where
\begin{eqnarray}\label{ProbTerms}
\mathcal{P}_c&=&\frac{1}{2}\cos ^{2}\alpha \cos ^{2}\frac{\phi _{1}}{2}+\frac{1}{4} \sin ^{2}\alpha,\nonumber\\
\mathcal{P}_s&=&\frac{1}{2}\cos ^{2}\alpha \sin ^{2}\frac{\phi _{1}}{2}+\frac{1}{4} \sin ^{2}\alpha,\nonumber\\
\mathcal{I}_c&=&\frac{1}{2\sqrt{2}}\sin 2\alpha \cos ^{2}\frac{\phi _{1}}{2},\nonumber\\
\mathcal{I}_s&=&\frac{1}{2\sqrt{2}}\sin 2\alpha \sin \frac{\phi _{1}}{2} \sin\left(\frac{\phi _{1}}{2}-\phi _{2}\right).
\end{eqnarray}
We remark that the terms $\mathcal{I}_c$, $\mathcal{I}_s$ in the detection probabilities exclusively stem from the interference between the $\ket{\mathrm{wave}}$ and $\ket{\mathrm{particle}}$ components appearing in the generated superposition state $\left| \psi _{f}\right\rangle$ of equation~(\ref{wp-state}). This fact is further evidenced by the appearance, in these interference terms, of the factor $\mathcal{C}=\sin2\alpha$, which is the amount of quantum coherence owned by $\left| \psi _{f}\right\rangle$ in the basis 
$\{\ket{\mathrm{wave}}, \ket{\mathrm{particle}}\}$ \cite{adessoReview}. 
On the other hand, the interference terms $\mathcal{I}_c$, $\mathcal{I}_s$ are always zero when the final state of the photon is: (i) $\ket{\mathrm{wave}}$ ($\alpha = 0$); (ii) $\ket{\mathrm{particle}}$ ($\alpha = \pi/2$); (iii) a classical incoherent mixture $\rho_{f}=\cos ^{2}\alpha \left| \mathrm{wave}\right\rangle\left\langle \mathrm{wave} \right| +\sin ^{2}\alpha \left| \mathrm{particle}\right\rangle \left\langle \mathrm{particle}\right| $ (resulting from an initial mixed polarization state of the photon).

\begin{figure*}[tbp]
	\centering
	\includegraphics[width=\textwidth]{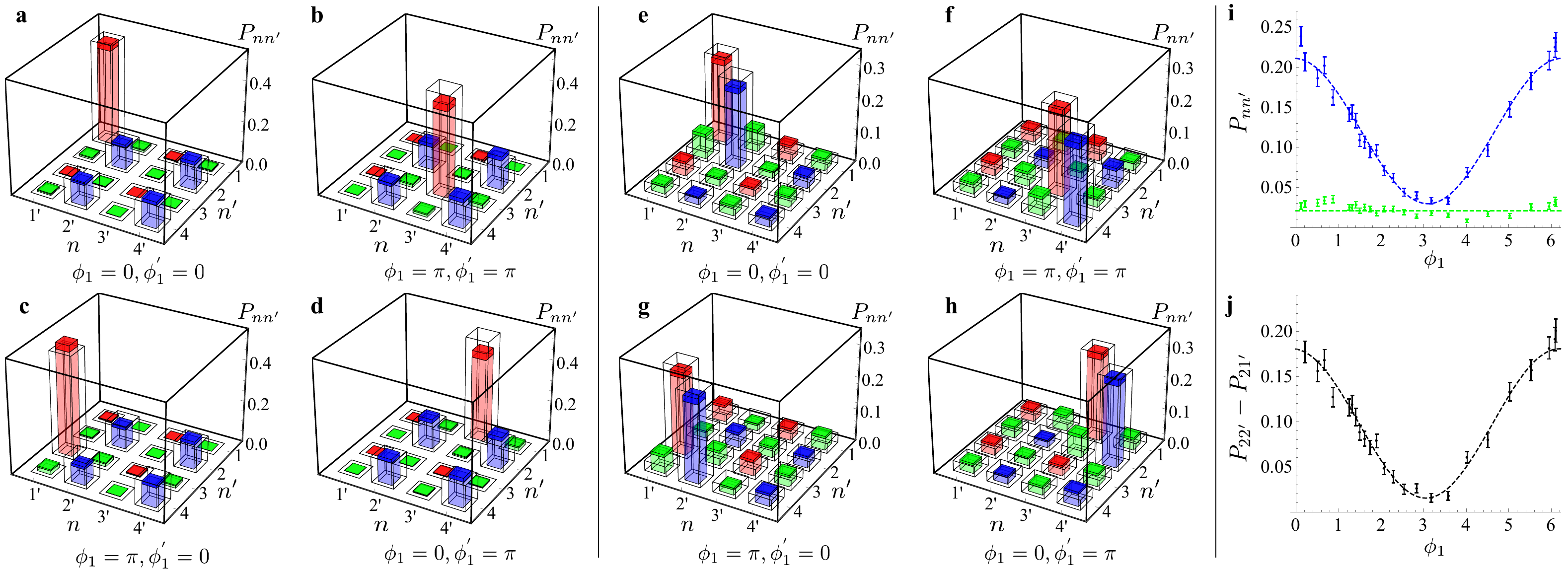}
	\caption{\textbf{ Generation of wave-particle entangled superposition with a two-photons state.} Measurements of the output coincidence probabilities  $P_{nn^{\prime }}$ to detect one photon in output mode $n$ of the first toolbox and one in the output mode $n^{\prime }$ of the second toolbox, with different phases $\phi_{1}$ and $\phi_{1}^{\prime }$ ($\phi_{2}=\phi_{2}^{\prime}=0$). 
	\textbf{(a-d)}, $P_{nn'}$ measured with $\{ \beta=0, \beta^{\prime} = 0 \}$, corresponding to the absence of BS$_4$ and BS$_5$ in Fig. \ref{fig:figure1}. {\bf a}, $\phi_{1}=0$ and $\phi_{1}^{\prime }=0$. {\bf b}, $\phi_{1}=\pi$ and $\phi_{1}^{\prime }=\pi$.  {\bf c},  $\phi_{1}=\pi$ and $\phi_{1}^{\prime }=0$.  {\bf d}, $\phi_{1}=\pi$ and $\phi_{1}^{\prime }=\pi$. 
	\textbf{(e-h)}, $P_{nn'}$ measured with $\{ \beta=22.5^{\circ}, \beta^{\prime}=22.5^{\circ} \}$, corresponding to the presence of BS$_4$ and BS$_5$ in Fig. \ref{fig:figure1}. {\bf e}, $\phi_{1}=0$ and $\phi_{1}^{\prime }=0$. {\bf f}, $\phi_{1}=\pi$ and $\phi_{1}^{\prime }=\pi$.  {\bf g},  $\phi_{1}=\pi$ and $\phi_{1}^{\prime }=0$.  {\bf h}, $\phi_{1}=\pi$ and $\phi_{1}^{\prime }=\pi$. 
	White boxes: theoretical predictions. Colored boxes: experimental data. Darker regions correspond to 1 $\sigma$ error, due to the Poissonian statistics of two-photon coincidences.
Red boxes highlight the detectors linked to wave-like behavior for both photons at $\{ \beta=0, \beta^{\prime} = 0 \}$. Blue boxes highlight the detectors linked to particle-like behavior for both photons at $\{ \beta=0, \beta^{\prime} = 0 \}$. Green boxes highlight the detectors linked to wave-like behavior for one photon and particle-like behavior for the other one at $\{ \beta=0, \beta^{\prime} = 0 \}$.
	\textbf{(i-j)}, Measurements of wave-particle entanglement witness. {\bf i}, $P_{22^{\prime}}$ (blue) and $P_{21^{\prime}}$ (green) and {\bf j}, witness $\mathcal{W}_E = P_{22^{\prime}} - P_{21^{\prime}}$, as a function of $\phi_{1}$ for $\phi_{1}^{\prime}=0$ and $\{ \beta=22.5^{\circ}, \beta^{\prime}=22.5^{\circ} \}$. 
		Error bars are due to the Poissonian statistics of two-photon coincidences. Dashed curves: best-fit of the experimental data.}
		\label{fig:figure4}
\end{figure*}

The experimental single-photon toolbox, realizing the proposed scheme of Fig.~\ref{fig:figure1}, is displayed in Fig.~\ref{fig:figure2}. The implemented layout presents the advantage of being interferometrically stable, thus not requiring active phase stabilization between the modes.
For $\alpha =0$, the photon is vertically polarized and entirely reflected from the PBS to
travel along path $1$, then split at BS$_{1}$ into two paths, both leading to the same BS$_{3}$ which allows these two paths to interfere with each other before detection. The photon detection probability at each detector $D_{n}$ 
$(n=1,2,3,4)$ depends on the phase shift $\phi _{1}$: $P_{1}(\alpha
=0)=P_{2}(\alpha =0)=\frac{1}{2}\cos ^{2}\frac{\phi _{1}}{2}$, 
$P_{3}(\alpha =0)=P_{4}(\alpha =0)=\frac{1}{2}\sin ^{2}\frac{\phi _{1}}{2}$,
as expected from equations (\ref{Pn}) and (\ref{ProbTerms}). After many such runs an
interference pattern emerges, exhibiting the wave-like nature of the photon.
Differently, if initially $\alpha =\pi /2$, the photon is horizontally polarized and, as a
whole, transmitted by the PBS to path $2$, then split at BS$_{2}$
into two paths (leading, respectively, to BS$_{4}$ and BS$_{5}$) 
which do not interfere anywhere. Hence, the phase shift $\phi _{2}$ plays no role
on the photon detection probability and each detector has an equal chance to click: $P_{1}(\alpha =
\frac{\pi }{2})=P_{2}(\alpha =\frac{\pi }{2})=P_{3}(\alpha =\frac{\pi }{2}
)=P_{4}(\alpha =\frac{\pi }{2})=\frac{1}{4}$, as predicted by equations (\ref{Pn}) and (\ref{ProbTerms}), showing particle-like nature without any interference pattern. Interestingly, for $0<\alpha <\pi /2,$ the photon simultaneously behaves like wave and particle. The continuous morphing transition from wave to particle behavior as $\alpha$ varies from $0$ to $\pi /2$ is clearly seen from Fig.~\ref{fig:figure3}c. The coherence witness defined as $\mathcal{W}_C = |P_{1} - P_{2}|$ is also measured. According to equations~(\ref{Pn}) and (\ref{ProbTerms}), $\mathcal{W}_C = 2\mathcal{I}_c$ is zero if and only if there is no wave-particle coherence. As shown in Fig.~\ref{fig:figure3}c, $\mathcal{W}_C$ testifies for both coherent $\ket{\psi_f}$ and mixed $\rho_f$ wave-particle states (the latter being obtained by adding a relative time delay in the interferometer paths larger than the photon coherence time to lose quantum interference).

\textbf{Wave-particle entanglement.} The above single-photon scheme constitutes the basic toolbox which can be extended to create the wave-particle entangled state of two photons, as shown in Fig. \ref{fig:figure2}b. Initially, a two-photon
polarization maximally entangled state 
$\left| \Psi \right\rangle _{AB}=\frac{1}{\sqrt{2}} (\left| VV\right\rangle
+\left| HH\right\rangle)$ is prepared (the procedure works in general for arbitrary weights, see Supplemental Information). Each photon is then sent to one of two identical wave-particle toolboxes which provide the final state
\begin{equation} \label{ent}
\left| \Phi \right\rangle _{AB}=\frac{1}{\sqrt{2}} \left(\left| \mathrm{wave} \right\rangle \left| \mathrm{wave}^{\prime }\right\rangle 
+ \left| \mathrm{particle}\right\rangle \left| \mathrm{particle}^{\prime}\right\rangle\right), 
\end{equation}
where the single-photon states $\left|\mathrm{wave}\right\rangle$, $\left|\mathrm{particle}\right\rangle$, 
$\left|\mathrm{wave}^{\prime}\right\rangle$, $\left| \mathrm{particle}^{\prime}\right\rangle$ are defined in equation~(\ref{wave-particle}), with parameters and paths related to the corresponding wave-particle toolbox.
The generated state $\left| \Phi \right\rangle _{AB}$ is thus a wave-particle maximally entangled state (Bell state) of two photons in separated locations. 

The output two-photon state is measured after the two toolboxes. The results are shown in Fig. \ref{fig:figure4}. Coincidences between the four outputs of each toolbox are measured by varying $\phi_{1}$ and $\phi_{1}^{\prime}$. The first set of measurements (Fig.~\ref{fig:figure4}a-d) is performed by setting the angles of the output wave-plates (see Fig.~\ref{fig:figure2}c) at $\{ \beta=0, \beta'=0\}$, corresponding to removing both BS$_{4}$ and BS$_{5}$ in Fig.~\ref{fig:figure1} (absence of interference between single-photon wave-like and particle-like behaviors). In this case, detectors placed at outputs (1,3) and (1$^{\prime}$,3$^{\prime}$) reveal wave-like behavior, while detectors placed at outputs (2,4) and (2$^{\prime}$,4$^{\prime}$) evidence a particle-like one. As expected, the two-photon probabilities $P_{nn^{\prime}}$ for the particle detectors remain unchanged while varying $\phi_{1}$ and $\phi_{1}^{\prime}$, whereas the $P_{nn^{\prime}}$ for the wave detectors show interference fringes. Moreover, no contribution of crossed (wave-like)-(particle-like) coincidences $P_{nn^{\prime}}$ is obtained, due to the form of the entangled state.
The second set of measurements (Fig. \ref{fig:figure4}e-h) is performed by setting the angles of the output wave-plates at angles 
$\{\beta=22.5^{\circ}, \beta'=22.5^{\circ}\}$, corresponding to the presence of BS$_{4}$ and BS$_{5}$ in Fig. \ref{fig:figure1} (presence of interference between single-photon wave and particle behaviors). We now observe nonzero contributions across all the probabilities depending on the specific settings of phases $\phi_{1}$ and $\phi_{1}^{\prime}$. In order to assess the presence of entanglement in the wave-particle natures, we also measure the wave-particle entanglement witness defined as $\mathcal{W}_E = P_{22'} - P_{21'}$ by varying $\phi_{1}$ with fixed $\phi_{1}^{\prime }=\phi_{2}=\phi_{2}^{\prime }=0$. According to the general expressions of the coincidence probabilities (see Supplemental Information), $\mathcal{W}_E$ is identically zero if and only if the wave-particle two-photon state is separable (e.g., $\ket{\mathrm{wave}}\otimes\ket{\mathrm{wave'}}$ or a maximal mixture of two-photon wave and particle states). For 
$\ket{\Phi}_{AB}$ of equation~(\ref{ent}) the theoretical prediction is $\mathcal{W}_E=(1/4)\cos^2(\phi_1/2)$, which is confirmed by the results reported in Fig.~\ref{fig:figure4}i-j (within the reduction due to visibility). These observations altogether prove the expected quantum correlations between wave and particle states of two photons in the entangled state $\ket{\Phi}_{AB}$.

\textbf{Conclusions.} In summary, we have introduced and realized a novel all-optical scheme to deterministically generate single-photon wave-particle superposition states. This setup has enabled the observation of the simultaneous coexistence of particle and wave character of the photon maintaining all its devices fixed, being the control only on the preparation of the input photon. Specifically, different initial polarization states of the photon, then transformed into which-way (path) states, reveal the wave-to-particle morphing economizing the employed resources compared to previous experiments with delayed choice \cite{quan-delay-exp,delay-exp1,delay-exp2,delay-exp3,delay-exp4,delay-exp5,pho-test}. 
The advantageous aspects of the single-photon scheme have then supplied the key for its straightforward doubling, by which we have observed that two photons can be cast in a wave-particle entangled state provided that suitable initial polarization entangled states are injected into the apparatus. We remark that powerful features of the scheme are flexibility and scalability. 
Indeed, a parallel assembly of $N$ single-photon wave-particle toolboxes allows the generation of $N$-photon wave-particle entangled states. For instance, the GHZ-like state $\ket{\Phi_N}=\frac{1}{\sqrt{2}} (\ket{\mathrm{wave}_1,\mathrm{wave}_2,\ldots,\mathrm{wave}_N}+\ket{\mathrm{particle}_1,\mathrm{particle}_2,\ldots,\mathrm{particle}_N})$ is produced when the GHZ polarization entangled state $\ket{\Psi_N}=\frac{1}{\sqrt{2}} (\ket{V_1V_2\ldots V_N}+\ket{H_1H_2\ldots H_N})$ is used as input state.

From a fundamental viewpoint, our research brings the complementarity principle for wave-particle duality to a further level. In fact, besides confirming that a photon can live in a superposition of wave and particle behaviors when observed by quantum detection \cite{delay-exp5}, we prove that the manifestation of its dual nature can intrinsically depend on the character of another photon, according to correlations ruled by quantum entanglement. In this case, the wave-particle behavior of a photon is determined by a measurement apparatus placed in a region spatially separated from it. This new phenomenon, merging complementarity principle and entanglement, can be named ``wave-particle duality action at a distance''.  
We finally highlight that the possibility to create and control wave-particle entanglement may also play a role in quantum information scenarios. In particular, it opens the way to design protocols which exploit quantum resources contained in systems of qubits encoded in wave and particle operational states.

\section*{Acknowledgements}
This work was supported by the ERC-Starting Grant 3D-QUEST (3D-Quantum Integrated Optical Simulation; grant agreement no. 307783, http://www.3dquest.eu) and by the Marie Curie Initial Training Network PICQUE (Photonic Integrated Compound Quantum Encoding, grant agreement no. 608062, funding Program: FP7-PEOPLE-2013-ITN, http://www.picque.eu).
In this work Z.X.M. and Y.J.X. are supported by National Natural Science Foundation of China under Grant no. 11574178 and no. 61675115, Shandong Provincial Natural Science Foundation, China under Grant no. ZR2016JL005, while N.B.A. is funded by the Vietnam National Foundation for Science and Technology Development (NAFOSTED) under project no. 103.01-2017.08.

\end{document}


\title{Supplemental Information\\ 
Entanglement of photons in their dual wave-particle nature}

\author{Adil S. Rab}
\affiliation{Dipartimento di Fisica, Sapienza Universit\`{a} di Roma, Piazzale Aldo Moro, 5, I-00185 Roma, Italy}

\author{Emanuele Polino}
\affiliation{Dipartimento di Fisica, Sapienza Universit\`{a} di Roma, Piazzale Aldo Moro, 5, I-00185 Roma, Italy}

\author{Zhong-Xiao Man} 
\email{zxman@mail.qfnu.edu.cn}
\affiliation{Shandong Provincial Key Laboratory of Laser Polarization and Information Technology, Department of Physics, Qufu Normal University, Qufu 273165, China}

\author{Nguyen Ba An}
\affiliation{Center for Theoretical Physics, Institute of Physics, Vietnam Academy of Science and Technology (VAST), 18
Hoang Quoc Viet, Cau Giay, Hanoi, Vietnam}

\author{Yun-Jie Xia}
\affiliation{Shandong Provincial Key Laboratory of Laser Polarization and Information Technology, Department of Physics, Qufu Normal University, Qufu 273165, China}

\author{Nicol\`{o} Spagnolo}
\affiliation{Dipartimento di Fisica, Sapienza Universit\`{a} di Roma, Piazzale Aldo Moro, 5, I-00185 Roma, Italy}

\author{Rosario Lo Franco} 
\email{rosario.lofranco@unipa.it}
\affiliation{Dipartimento di Energia, Ingegneria dell'Informazione e Modelli Matematici, Universit\`{a} di Palermo, Viale delle Scienze, Edificio 9, 90128 Palermo, Italy}
\affiliation{Dipartimento di Fisica e Chimica, Universit\`a di Palermo, via Archirafi 36, 90123 Palermo, Italy}

\author{Fabio Sciarrino}
\email{fabio.sciarrino@uniroma1.it}
\affiliation{Dipartimento di Fisica, Sapienza Universit\`{a} di Roma, Piazzale Aldo Moro, 5, I-00185 Roma, Italy}

\maketitle

\section*{SUPPLEMENTARY NOTE 1: SINGLE-PHOTON WAVE-PARTICLE STATE}

In this Section we describe the derivation of the wave-particle superposition state of a single photon which travels along the theoretical scheme, reported in Fig.~1 of the main text (see also Supplementary Fig.~\ref{fig:figureS1} below in Supplementary Note 3) and realized by the experimental setup (wave-particle toolbox) of Fig.~2 of the manuscript.

A photon is initially prepared in a polarization state
\begin{equation}
\left| \psi _{0}\right\rangle =\cos \alpha \left| V\right\rangle +\sin
\alpha \left| H\right\rangle ,
\end{equation}
with $\left| V\right\rangle $ and $\left| H\right\rangle $ representing the
states of vertical and horizontal polarization, respectively. This state is experimentally realized by a half-wave plate (HWP) not shown in Fig.~1 (but evidenced in Fig.~2a in the manuscript).
The photon so prepared is sent to a polarizing beam splitter (PBS). Since the vertical polarization of the photon is reflected by the PBS (path 1), while horizontal polarization is transmitted through the PBS (path 2), the photon passes through the upper (wave-like) path of Fig.~1 with a probability amplitude $\cos\alpha $ and it crosses the lower (particle-like) path with a probability 
amplitude $\sin \alpha$. A HWP ($45\degree$) is placed after the PBS to obtain equal polarizations between the two spatial modes (paths). Therefore, after the PBS and HWP, the photon state is
\begin{equation}
\left| \psi _{1}\right\rangle =\cos \alpha \left| 1\right\rangle +\sin
\alpha \left| 2\right\rangle ,
\end{equation}
where $\left| n\right\rangle $ ($n=1,2,3,4$) represents a state of a photon
traveling along path $n$. Then, each path further bifurcates at a balanced
beam splitter (BS), BS$_{1}$ for path $1$ and BS$_{2}$ for path $2$, transforming $\left| \psi _{1}\right\rangle $ into
\begin{equation}
\left| \psi _{2}\right\rangle =\cos \alpha \left[\frac{1}{\sqrt{2}}(\left|1\right\rangle +e^{i\phi _{1}}\left| 3\right\rangle )\right]
+\sin \alpha \left[\frac{1}{\sqrt{2}}(\left| 2\right\rangle +e^{i\phi _{2}}\left| 4\right\rangle )\right],
\end{equation}
where $\phi _{1}$ ($\phi _{2}$) is a relative phase introduced by a phase shifter placed in path $3$ ($4$). Notice that paths 3 and 4 are the paths reflected by BS$_1$ and BS$_2$, respectively. 
Paths 1 and 3 are then recombined by BS$_{3}$, after which the state $\left|
\psi _{2}\right\rangle$ becomes
\begin{equation}
\left| \psi _{3}\right\rangle =e^{i\phi _{1}/2}\cos \alpha \left[\cos \frac{\phi
_{1}}{2}\left| 1\right\rangle -i\sin \frac{\phi _{1}}{2}\left|
3\right\rangle \right]+\sin \alpha \left[\frac{1}{\sqrt{2}}(\left| 2\right\rangle
+e^{i\phi _{2}}\left| 4\right\rangle )\right].
\end{equation}
We remark that, already at this stage (that is, without BS$_4$ and BS$_5$ in Fig.~1), the photon state is a superposition of a wave-like state ($e^{i\phi _{1}/2} [\cos \frac{\phi_{1}}{2}\left| 1\right\rangle -i\sin \frac{\phi _{1}}{2}\left|3\right\rangle]$) and a particle-like state ($\frac{1}{\sqrt{2}}[\left| 2\right\rangle+e^{i\phi _{2}}\left| 4\right\rangle]$). Photon counting probabilities at detectors $D_1$, $D_3$ placed at the end of paths 1, 3 will reveal a wave-like behavior with their dependence on the phase $\phi_1$; on the other hand, photon counting probabilities at detectors $D_2$, $D_4$ placed at the end of paths 2, 4 will exhibit a particle-like behavior independent of the phase $\phi_2$ (see Fig. 4a-d in the main text). However, these counting probabilities do not allow us to observe a wave-to-particle morphing since there is no interference between the wave-like and particle-like states at the detection level. It is like measuring the state $\left| \psi _{3}\right\rangle$ only along the orthogonal basis corresponding to wave-like or particle-like behavior.

In order to observe a wave-to-particle morphing as a function of the parameter $\alpha$, the wave-like and particle-like behaviors have to interfere at the detection level. This is achieved by letting paths $1$ and $2$ synchronize at BS$_{4}$, while paths $3$ and $4$ synchronize at BS$_{5}$. This way, the state $\ket{\psi_3}$ is turned into the final state of the photon 
\begin{equation}
\left| \psi _{f}\right\rangle =\cos \alpha \left| \mathrm{wave}\right\rangle
+\sin \alpha \left| \mathrm{particle}\right\rangle ,  \label{wp-state}
\end{equation}
where
\begin{equation}
\left| \mathrm{wave}\right\rangle =\frac{1}{\sqrt{2}}e^{i\phi _{1}/2}(\cos
\frac{\phi _{1}}{2}\left| 1\right\rangle +\cos \frac{\phi _{1}}{2}\left|
2\right\rangle -i\sin \frac{\phi _{1}}{2}\left| 3\right\rangle -i\sin \frac{
\phi _{1}}{2}\left| 4\right\rangle ),  \label{wave}
\end{equation}
and
\begin{equation}
\left| \mathrm{particle}\right\rangle =\frac{1}{2}(\left| 1\right\rangle
-\left| 2\right\rangle +e^{i\phi _{2}}\left| 3\right\rangle -e^{i\phi
_{2}}\left| 4\right\rangle ),  \label{particle}
\end{equation}
statistically describe wave and particle behavior of the photon, respectively. Now, as explicitly reported in the main text, photon counting probabilities at detectors $D_n$ ($n=1,2,3,4$) reveal the quantum superposition of wave and particle behaviors and the wave-to-particle morphing. The inception of BS$_{4}$ and BS$_{5}$ in fact permits one to change the measurement basis into the coherent superposition of wave-like and particle-like behaviors. In the manuscript, we notice that the terms $\mathcal{I}_c$, $\mathcal{I}_s$ of the detection probabilities exclusively stem from the interference between the $\ket{\mathrm{wave}}$ and $\ket{\mathrm{particle}}$ components appearing in the generated superposition state $\left| \psi _{f}\right\rangle$ of equation~(\ref{wp-state}). In these interference terms, the factor $\mathcal{C}=\sin2\alpha$ appears, which is the amount of quantum coherence owned by $\left| \psi _{f}\right\rangle$ in the basis $\{\ket{\mathrm{wave}}, \ket{\mathrm{particle}}\}$. This is obtained by using a bona-fide quantifier of quantum coherence for a two-state system, defined as $\mathcal{C}=\sum_{i,j\ (i\neq j)}|\rho_{ij}|$, where $\rho_{ij}$ are the off-diagonal terms of the system density matrix [S1]. 

It is then immediate to see that, starting from a mixed polarization photon state of the kind $\rho_{0}=\cos ^{2}\alpha \left| V\right\rangle\left\langle V \right| +\sin ^{2}\alpha \left| H\right\rangle \left\langle H\right| $, a mixed wave-particle state for the photon is finally obtained by the scheme above, namely: $\rho_{f}=\cos ^{2}\alpha \left| \mathrm{wave}\right\rangle\left\langle \mathrm{wave} \right| +\sin ^{2}\alpha \left| \mathrm{particle}\right\rangle \left\langle \mathrm{particle}\right| $.

\section*{SUPPLEMENTARY NOTE 2: TWO-PHOTON WAVE-PARTICLE ENTANGLED STATE}

We now describe the steps leading to the generation of the wave-particle entangled state of two separated photons. The scheme is a parallel doubling of the single-photon scheme of Fig.~1 (see Fig.~2 of the manuscript). In order to give the most general theoretical description of the procedure, let us consider the injection of an initial polarization entangled state of the form
\begin{equation}
\left| \Psi \right\rangle _{AB}=\cos \alpha \left| VV\right\rangle
_{AB}+\sin \alpha \left| HH\right\rangle _{AB}.  \label{3}
\end{equation}
Then, the photon $A$ ($B$) is sent to wave-particle toolbox $A$, top, ($B$, bottom) as displayed in Fig.~2b of the manuscript. 
The photons therefore independently follow the same steps described in the section above. For simplicity, we indicate the parameters, optical devices and paths of the bottom wave-particle toolbox ($B$) with the symbol ($^\prime$).

As a consequence, after the PBS + HWP and PBS$'$ + HWP$'$, the two-photon state becomes
\begin{equation}
\left| \Psi _{1}\right\rangle =\cos \alpha \left| 11'\right\rangle +\sin
\alpha \left| 22'\right\rangle ,
\end{equation}
where we indicate $\ket{nn'}\equiv \ket{n}\ket{n'} = \ket{n}\otimes\ket{n'}$.

Then, paths $1$, $1'$ goes towards BS$_1$, BS$_{1'}$, while paths $2$, $2'$ bifurcate at BS$_2$, BS$_{2'}$, transforming $\left| \Psi _{1}\right\rangle$ into
\begin{equation}
\left| \Psi _{2}\right\rangle =\cos \alpha \left[\frac{1}{\sqrt{2}}(\left| 1\right\rangle +e^{i\phi _{1}}\left| 3\right\rangle )\right]
\left[\frac{1}{\sqrt{2}}(\left|1'\right\rangle +e^{i\phi _{1}'}\left| 3'\right\rangle )\right]
+\sin \alpha\left[\frac{1}{\sqrt{2}}(\left| 2\right\rangle +e^{i\phi _{2}}\left| 4\right\rangle)\right]
\left[\frac{1}{\sqrt{2}}(\left| 2'\right\rangle +e^{i\phi _{2}'}\left| 4'\right\rangle )\right],
\end{equation}
where $\phi _{1}$, $\phi _{1}'$, $\phi _{2}$ and $\phi _{2}'$\ are the relative phases
introduced by the phase shifters placed in path $3$, $3'$, $4$ and $4'$.
Successively, paths 1 and 3 are recombined by BS$_{3}$ and paths $1'$ and $3'$ are recombined by BS$_{3'}$,
after which the state $\left|\Psi _{2}\right\rangle $ becomes
\begin{eqnarray}\label{EntPsi3}
\ket{\Psi _{3}} &=& \cos\alpha \left[e^{i\phi_{1}/2} \left(\cos \frac{\phi_{1}}{2}\left| 1\right\rangle -i\sin \frac{\phi _{1}}{2}\left|3\right\rangle\right)\right] \left[e^{i\phi'_{1}/2} \left(\cos \frac{\phi'_{1}}{2}\left| 1'\right\rangle -i\sin \frac{\phi _{1}'}{2}\left|3'\right\rangle\right)\right] \nonumber\\
&+& \sin\alpha \left[\frac{1}{\sqrt{2}}\left(\left| 2\right\rangle+e^{i\phi _{2}}\left| 4\right\rangle\right)\right] \left[\frac{1}{\sqrt{2}}\left(\left| 2'\right\rangle+e^{i\phi' _{2}}\left| 4'\right\rangle\right)\right].
\end{eqnarray}
At this stage, that is removing the final beam splitters BS$_4$, BS$_5$ and BS$_{4'}$, BS$_{5'}$ in each wave-particle toolbox, the entangled state can be measured by photon coincidences $P_{nn'}$ detecting their wave-like or particle-like behaviors.
In the manuscript, we have performed such a measurement for the case when the two-photon state is maximally entangled (that is, $\alpha=\pi/4$ in equation~(\ref{EntPsi3})). The theoretical probabilities corresponding to wave-like, particle-like and crossed (wave-like)-(particle-like) behaviors are as follows. 
Wave-like probabilities:
\begin{eqnarray}
P_{11^{\prime}}&=&\frac{1}{2}\cos^2 \frac{\phi_{1}}{2}\cos^2 \frac{\phi_{1}^{\prime }}{2},\quad
P_{33^{\prime}}=\frac{1}{2}\sin^2 \frac{\phi_{1}}{2}\sin^2 \frac{\phi_{1}^{\prime }}{2}, \nonumber\\
P_{13^{\prime}}&=&\frac{1}{2}\cos^2 \frac{\phi_{1}}{2}\sin^2 \frac{\phi_{1}^{\prime }}{2},\quad
P_{31^{\prime}}=\frac{1}{2}\sin^2 \frac{\phi_{1}}{2}\cos^2 \frac{\phi_{1}^{\prime }}{2}.
\end{eqnarray}
Particle-like probabilities:
\begin{equation}
P_{22^{\prime}}=P_{44^{\prime}}=P_{24^{\prime}}=P_{42^{\prime}}=1/8.
\end{equation}
Crossed (wave-like)-(particle-like) probabilities:
\begin{equation}
P_{12^{\prime}}=
P_{14^{\prime}}=
P_{21^{\prime}}=
P_{23^{\prime}}=
P_{32^{\prime}}=
P_{34^{\prime}}=
P_{41^{\prime}}=
P_{43^{\prime}}=0.
\end{equation} 
The experimental results for these probabilities are plotted in Fig.~4a-d of the manuscript.

When the final beam splitters in each wave-particle toolbox are used, paths $1$ ($1'$) and $2$ ($2'$) interfere at BS$_{4}$\ (BS$_{4'}$) while paths $3$ ($3'$) and $4$ ($4'$) synchronize at BS$_{5}$ (BS$_{5'}$). The final state of the two photons is
\begin{equation}\label{entfinal}
\left| \Phi \right\rangle _{AB}=\cos \alpha \left| \mathrm{wave}
\right\rangle \left| \mathrm{wave}^{\prime }\right\rangle 
+\sin \alpha \left| \mathrm{particle}\right\rangle \left| \mathrm{particle}^{\prime }\right\rangle, 
\end{equation}
where the states $\left| \mathrm{wave}\right\rangle$ and $\left|\mathrm{particle}\right\rangle$ are defined in equations~(\ref{wave}) and (\ref{particle}), while the states $\left| \mathrm{wave}^{\prime
}\right\rangle$ and $\left| \mathrm{particle}^{\prime
}\right\rangle $ are defined in the same way, namely
\begin{eqnarray}
\left| \mathrm{wave}^{\prime }\right\rangle &=& \frac{1}{\sqrt{2}}
e^{i\phi _{1}^{\prime }/2}(\cos \frac{\phi _{1}^{\prime }}{2}\left|
1^{\prime }\right\rangle+\cos \frac{\phi _{1}^{\prime }}{2} \left| 2^{\prime
}\right\rangle -i\sin \frac{\phi _{1}^{\prime }}{2}\left| 3^{\prime
}\right\rangle -i\sin \frac{\phi _{1}^{\prime }}{2}\left| 4^{\prime
}\right\rangle )\nonumber,\\
\left| \mathrm{particle}^{\prime }\right\rangle &=&\frac{1}{2}
(\left| 1^{\prime }\right\rangle -\left| 2^{\prime}\right\rangle +e^{i\phi
_{2}^{\prime }}\left| 3^{\prime }\right\rangle -e^{i\phi _{2}^{\prime
}}\left| 4^{\prime }\right\rangle ).
\end{eqnarray}
For the state $\ket{\Phi}_{AB}$, the coincidence probabilities $P_{nn^{\prime }}=P_{nn^{\prime }}(\alpha ,\phi _{1},\phi
_{1}^{\prime },\phi _{2},\phi _{2}^{\prime })$ that a pair of detectors $
D_{n}$\ and $D_{n^{\prime }}$\ $\left( n=1,2,3,4;\text{ }n^{\prime
}=1^{\prime },2^{\prime },3^{\prime },4^{\prime }\right) $ fires are found to be
\begin{eqnarray}\label{TwophotonProb}
P_{11^{\prime }} &=&P_{22^{\prime }}=\frac{1}{4}\cos ^{2}\alpha \cos ^{2}\frac{\phi _{1}}{2}%
\cos ^{2}\frac{\phi _{1}^{\prime }}{2}+\frac{1}{16}\sin ^{2}\alpha +\frac{1}{%
8}\sin 2\alpha \cos \frac{\phi _{1}}{2}\cos \frac{\phi _{1}^{\prime }}{2}%
\cos \left(\frac{\phi _{1}+\phi _{1}^{\prime }}{2}\right), \nonumber \\
P_{12^{\prime }} &=&P_{21^{\prime }}=\frac{1}{4}\cos ^{2}\alpha \cos ^{2}\frac{\phi _{1}}{2}%
\cos ^{2}\frac{\phi _{1}^{\prime }}{2}+\frac{1}{16}\sin ^{2}\alpha -\frac{1}{%
8}\sin 2\alpha \cos \frac{\phi _{1}}{2}\cos \frac{\phi _{1}^{\prime }}{2}%
\cos \left(\frac{\phi _{1}+\phi _{1}^{\prime }}{2}\right),  \nonumber \\
P_{13^{\prime }} &=&P_{24^{\prime }}=\frac{1}{4}\cos ^{2}\alpha \cos ^{2}\frac{\phi _{1}}{2}%
\sin ^{2}\frac{\phi _{1}^{\prime }}{2}+\frac{1}{16}\sin ^{2}\alpha -\frac{1}{%
8}\sin 2\alpha \cos \frac{\phi _{1}}{2}\sin \frac{\phi _{1}^{\prime }}{2}%
\sin \left(\phi _{2}^{\prime }-\frac{\phi _{1}+\phi _{1}^{\prime }}{2}\right),
\nonumber \\
P_{14^{\prime }} &=&P_{23^{\prime }}=\frac{1}{4}\cos ^{2}\alpha \cos ^{2}\frac{\phi _{1}}{2}%
\sin ^{2}\frac{\phi _{1}^{\prime }}{2}+\frac{1}{16}\sin ^{2}\alpha+ \frac{1}{%
8}\sin 2\alpha \cos \frac{\phi _{1}}{2}\sin \frac{\phi _{1}^{\prime }}{2}%
\sin \left(\phi _{2}^{\prime }-\frac{\phi _{1}+\phi _{1}^{\prime }}{2}\right),\nonumber\\
P_{31^{\prime }} &=&P_{42^{\prime }}=\frac{1}{4}\cos ^{2}\alpha \sin ^{2}\frac{\phi _{1}}{2}%
\cos ^{2}\frac{\phi _{1}^{\prime }}{2}+\frac{1}{16}\sin ^{2}\alpha- \frac{1}{%
8}\sin 2\alpha \sin \frac{\phi _{1}}{2}\cos \frac{\phi _{1}^{\prime }}{2}%
\sin \left(\phi _{2}-\frac{\phi _{1}+\phi _{1}^{\prime }}{2}\right),\nonumber\\
P_{32^{\prime }} &=&P_{41^{\prime }}=\frac{1}{4}\cos ^{2}\alpha \sin ^{2}\frac{\phi _{1}}{2}%
\cos ^{2}\frac{\phi _{1}^{\prime }}{2}+\frac{1}{16}\sin ^{2}\alpha+ \frac{1}{%
8}\sin 2\alpha \sin \frac{\phi _{1}}{2}\cos \frac{\phi _{1}^{\prime }}{2}%
\sin \left(\phi _{2}-\frac{\phi _{1}+\phi _{1}^{\prime }}{2}\right),\nonumber\\
P_{33^{\prime }} &=&P_{44^{\prime }}=\frac{1}{4}\cos ^{2}\alpha \sin ^{2}\frac{\phi _{1}}{2}%
\sin ^{2}\frac{\phi _{1}^{\prime }}{2}+\frac{1}{16}\sin ^{2}\alpha- \frac{1}{
8}\sin 2\alpha \sin \frac{\phi _{1}}{2}\sin \frac{\phi _{1}^{\prime }}{2}
\cos \left(\phi _{2}+\phi _{2}^{\prime }-\frac{\phi _{1}+\phi _{1}^{\prime }}{2}\right),\nonumber\\
P_{34^{\prime }} &=&P_{43^{\prime }}=\frac{1}{4}\cos ^{2}\alpha \sin ^{2}\frac{\phi _{1}}{2}
\sin ^{2}\frac{\phi _{1}^{\prime }}{2}+\frac{1}{16}\sin ^{2}\alpha+ \frac{1}{
8}\sin 2\alpha \sin \frac{\phi _{1}}{2}\sin \frac{\phi _{1}^{\prime }}{2}
\cos \left(\phi _{2}+\phi _{2}^{\prime }-\frac{\phi _{1}+\phi _{1}^{\prime }}{2}\right).
\end{eqnarray}
The theoretical predictions of these sixteen probabilities for the generated maximally entangled state of the experiment reported in the manuscript, can be retrieved by fixing $\alpha=\pi/4$. The experimental plots are reported in Fig.~4e-h of the manuscript. We recall that the detection of these probabilities corresponds to measuring each photon of the entangled pair along the 
coherent superposition of wave-like and particle-like behaviors.

We point out that the third terms of all the coincidence probabilities of equation~(\ref{TwophotonProb}) are zero if and only if there is no quantum entanglement between the wave and particle degrees of freedom of the two photons. In fact, all these terms contain the factor $C=\sin 2\alpha$, where $C$ is the concurrence quantifying the entanglement of the state $\ket{\Phi}_{AB}$ of equation~(\ref{entfinal}) in the two-photon basis $\{\ket{\textrm{wave}}\ket{\textrm{wave$'$}},\ket{\textrm{wave}}\ket{\textrm{particle$'$}},\ket{\textrm{particle}}\ket{\textrm{wave$'$}},\ket{\textrm{particle}}\ket{\textrm{particle$'$}}\}$ [S2]. Therefore, for experimental aims, it is natural to introduce an entanglement witness defined for example as 
\begin{equation}
\mathcal{W}_E = P_{22'} - P_{21'} = \frac{1}{4}\sin 2\alpha \cos \frac{\phi _{1}}{2}\cos \frac{\phi _{1}^{\prime }}{2}
\cos \left(\frac{\phi _{1}+\phi _{1}^{\prime }}{2}\right),
\end{equation}
which is identically zero if and only if the wave-particle two-photon state is separable (unentangled): $\alpha=0$ ($\ket{\mathrm{wave}}\otimes\ket{\mathrm{wave'}}$), $\alpha=\pi/2$ ($\ket{\mathrm{particle}}\otimes\ket{\mathrm{particle'}}$). This witness would be zero also for a mixture of two-photon wave and particle states like $\rho_{AB}=\cos ^{2}\alpha \ket{\mathrm{wave}}\bra{\mathrm{wave}}\otimes \ket{\mathrm{wave'}}\bra{\mathrm{wave'}} + \sin ^{2}\alpha \ket{\mathrm{particle}}\bra{\mathrm{particle}}\otimes \ket{\mathrm{particle'}}\bra{\mathrm{particle'}}$, since the third terms in the probabilities do not appear at all. For the generated entangled state in the experiment, that is for $\ket{\Phi}_{AB}$ of equation~(\ref{entfinal}) with $\alpha=\pi/4$, fixing $\phi'_1=0$, this witness reduces to $\mathcal{W}_E=(1/4)\cos^2(\phi_1/2)$. Experimental measurements of the latter are reported in Fig.~4i-j of the manuscript.

Finally, we notice that by changing the initial polarization entangled state, different wave-particle entangled states can be created. For example, $\frac{1}{\sqrt{2}} (\left| VH\right\rangle +\left| HV\right\rangle)$ leads to $\frac{1}{\sqrt{2}} (\left| \mathrm{wave} \right\rangle \left| \mathrm{particle}^{\prime }\right\rangle + \left| \mathrm{particle}\right\rangle \left| \mathrm{wave}^{\prime}\right\rangle$). 

\section*{SUPPLEMENTARY NOTE 3: DESCRIPTION OF THE WAVE-PARTICLE TOOLBOX}

We now discuss the implementation of the wave-particle toolbox reported in Fig. 2c-d of the main text. The layout is shown in Supplementary Fig. \ref{fig:figureS1}b, while the conceptual scheme (Fig. 1 in the main text) is reported in Supplementary Fig. \ref{fig:figureS1}a. The implemented toolbox exploits simultaneously the polarization degree of freedom and different spatial modes in an interferometrically stable configuration.

\begin{suppFig}[ht!]
	\centering
	\includegraphics[width= 0.8\textwidth]{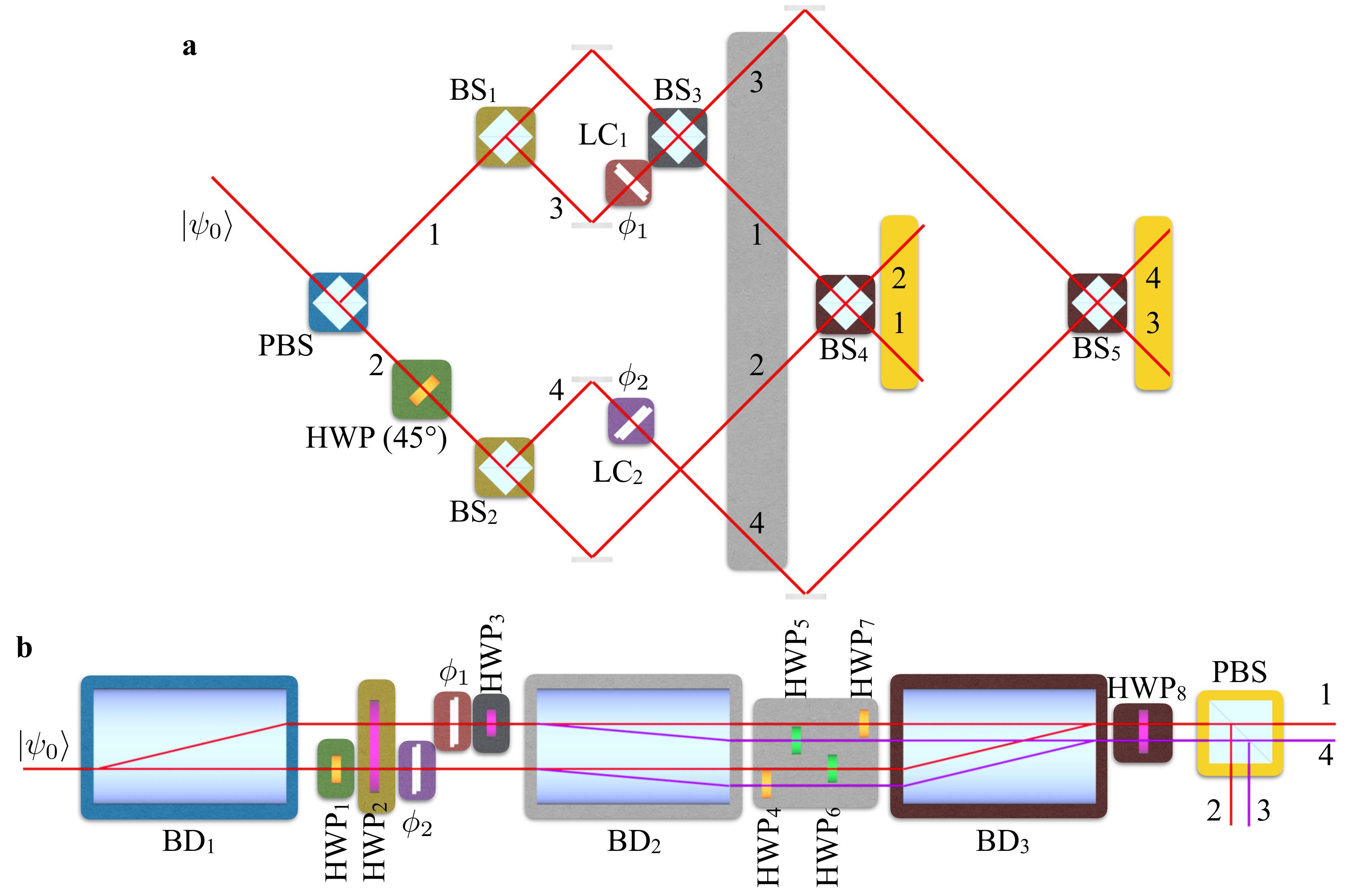} 
	\caption{{\bf a}, Conceptual scheme of the wave-particle toolbox. {\bf b}, Layout of the experimental implementation of the wave-particle toolbox. Angles of the optical axis orientations for the half-wave plates are: HWP$_1$($45^{\circ}$), HWP$_2$($22.5^{\circ}$), HWP$_3$($22.5^{\circ}$), HWP$_4$($45^{\circ}$), HWP$_5$($0^{\circ}$), HWP$_6$($0^{\circ}$), HWP$_7$($45^{\circ}$). Colors below the elements of the two panels identify the analogies between the two schemes.}
		\label{fig:figureS1}
\end{suppFig}

The input state $\vert \psi_{0} \rangle$ is separated in two parallel beams according to their polarization state by the first beam-displacing prism (BD$_1$), thus implementing the action of the first PBS. Then, HWP$_1$ with optical axis at $45^{\circ}$ is placed on the bottom mode of the interferometer. The second half-wave plate (HWP$_2$) implements simultaneously in the polarization degree of freedom the action of BS$_1$ and BS$_2$. After insertion of phases $\phi_{1}$ and $\phi_{2}$ between polarization states through liquid crystals LC$_1$ and LC$_2$, the action of BS$_3$ is reproduced by HWP$_3$ intercepting only the top mode. Then, beam-displacing prism BD$_2$ and the set of half-wave plates HWP$_4$-HWP$_7$ is inserted to separate and prepare the four output modes. Finally, the modes are recombined spatially by BD$_3$ and in polarization by HWP$_8$. Depending on the angle of HWP$_8$, this corresponds to removing ($\beta=0^{\circ}$) or inserting ($\beta=22.5^{\circ}$) the final beam-splitters BS$_4$ and BS$_5$ of the conceptual scheme. The final PBS in the experimental scheme of Supplementary Fig. \ref{fig:figureS1}b spatially separates the four output modes $1$-$4$, which are measured by detectors $D_1$-$D_4$. 

\section*{SUPPLEMENTARY REFERENCES}

\noindent [S1] Streisov, A., Adesso, G. \& Plenio, M. B. Quantum coherence as a resource. \emph{arXiv:1609.02439}.

\noindent [S2] Horodecki, R. Horodecki, P., Horodecki, M. \& Horodecki, K. Quantum entanglement. \emph{Rev. Mod. Phys.} \textbf{81}, 865--942 (2009).